# Evaluating the Wide Area Classroom After 24,000 HPC Students

John Urbanic
Pittsburgh Supercomputing Center
Carnegie Mellon University
Pittsburgh, PA, USA
E-mail: urbanic@psc.edu

Thomas Maiden
Pittsburgh Supercomputing Center
Carnegie Mellon University
Pittsburgh, PA USA
E-mail: tmaiden@psc.edu

Valerie Rossi
Pittsburgh Supercomputing Center
Carnegie Mellon University
Pittsburgh, PA, USA
E-mail: vrossi@psc.edu

## ABSTRACT

As of 2023 we have taught more than 24,000 students over the course of 106 events using the Wide Area Classroom, a novel distributed teaching platform. This has been a successful effort gauged by several important metrics. We describe both the technical and logistical structure of these events as well as the specific HPC curriculums which have proven to be most popular.

## CCS CONCEPT

- **Applied Computing → Education**

## 1 INTRODUCTION

The challenges of supporting the HPC training demands of the national scientific community are well understood [1], [2], and we have been grappling with them for decades as trainers at a national supercomputing facility. As requests for our training began to grow in conjunction with the growth of parallel computing (at the core, node, and now heterogenous levels) we found that our single-site workshop model was unable to keep up with demand.

This scarcity-of-expertise problem is widespread across education, and recent technological advancements have led to many improvement strategies ranging from YouTube tutorials to Massive Open Online Courses (MOOCs). While YouTube and Wikipedia have ushered in a golden age for autodidacts, the results have been less impressive for those wishing for a classroom experience. The early enthusiasm for MOOCs has given way to widespread reports of low satisfaction and dismal retention [3], [4]. The authors themselves have participated in such programs, and will continue to do so, although we now prefer our new alternative.

Small Private Online Courses (SPOCs) attempt to address these quality issues by scaling down while retaining some of the technological advances. However, with typical class sizes of less than two dozen students, this does not address the demand issue well. In addition, SPOCs also often have a high ratio of resources per student. This is acceptable in for-profit private training, but less so for limited academic budgets. We show below that we have been routinely successful with class sizes of 400 students, without compromising quality, and with costs similar to single-site events.

The longstanding practice at the Pittsburgh Supercomputing Center of collecting evaluation data after all training events gave us some insight into what ingredients were necessary for a useful experience within the HPC community. In particular, we came to the realization that classroom learning is a social process, and any method that discounts the ability of one student being able to turn to their fellow and ask, "Does this work for you?" is lacking a critical dimension. What instructor doesn't value subtle visual clues from the audience? Sticking a hand up in the middle of a lecture is the most effective way for a wayward student to keep themselves (and usually several classmates) from falling behind. And often the answer to that question becomes a dialog. Instant messaging and chat channel formats cannot duplicate this kind of interactivity.

Keeping these realities in mind, the team developed a scalable training approach that has resulted in the Wide Area Classroom (WAC). We wish to share the results of this effort with the HPC education and outreach community.

## 2 WIDE AREA CLASSROOM DESIGN

The Wide Area Classroom infrastructure has multiple important components that we are assembling into a "blue book" for others to use. It includes considerably detailed guidance, derived from our experience, so we will provide only the highlights in this section.

### 2.1 Audio-Visual Infrastructure

*2.1.1 Virtual Classroom Setup.* Our need to scale beyond a single site required the use of a telepresence of some sort. If we truly wanted to enable the "classroom" experience, with students being able to easily ask the instructor questions, it would require two-way HD video so that both the students and instructor could remain engaged. Our current solution is to have two distinct zoom feeds. One fully shows the instructor while the other is dedicated to the slide deck. This can provide redundancy through a well-known and universally accepted mechanism. The zoom chat allows the remote sites to be always in direct contact with the local TA and AV experts.

*2.1.2 Presentation Style.* The presenter has dual screens, allowing use of PowerPoint-type "presenter" mode - which provides a useful perspective on the presentation, and facilitates quick content navigation and virtual laser-pointer capabilities. The multiple screens permit high-resolution views of the remote sites, with any active site (asking a question) zoomed onto the rear screen directly in line-of-sight of the lecturer. Students at the production hub also have the materials projected onto the front screens. There are cameras for both archival recording as well as the live feed. Acoustics are important and have been arranged to provide good voice isolation for the speaker. The producer can monitor local and remote status to detect and resolve technical issues quickly. This is a particular arrangement, as indicated by some of the site-specific details, but could be replicated anywhere.

## 2.2 Class Sizing

*2.2.1 Sites.* A representative class might involve two dozen sites and as many as 400 students. This later number is no technical ceiling, but rather a requirement to maintain the classroom atmosphere. It could go no higher even if we could teleport all of our students to a single lecture hall. This is around the level of attendance where student questions tend to overwhelm the available time, at least for our typical HPC audiences and content. And we want to encourage live questions with a two-way dialog. There are chat-type back channels for the TAs to sort out very technical issues, or for code fragments and such, but these do not provide the kind of back-and-forth dialog that characterize the best of live instruction.

*2.2.2 Size Limitations.* Thus, this a sort of theoretical scaling limit of the WAC for highly technical material. We have participated - as instructors - in popular webinars, such as the *NVIDIA OpenACC Introduction and Advanced Courses*, which forgo these enrollment ceilings. Even with funding for excellent support personnel acting as TAs, the student interaction is greatly muted. For less technical material the ceiling could be raised, but if constant interaction and hands-on support are not critical to the event, then a webinar or MOOC model could suffice instead of the WAC.

*2.2.3 Sizing Cap.* The satellite site cap (currently around 25) is driven primarily by the limitations of the human monitor to deal with exceptional issues. It has grown as we have become more confident, but we remain conservative.

## 2.3 Support Personnel

We have mentioned TAs only briefly, but they are a critical part of the WAC formula. The site TAs do not need any domain expertise, but they do keep minor technical issues from becoming overwhelming at scale. They are often recruited from general IT support, but those that do have in-depth domain knowledge have contributed significantly. We have recently allowed a small number of TA-less sites for solitary individuals that have access or disability issues. They are listed in our data below as "Online" students. This has proved to be a valuable experiment. Although there is not yet enough data to report on online student satisfaction, it is clear that support issues make this a non-scalable mode of training.

## 2.4 Content Design

An easily overlooked pillar of the WAC is thoughtfully constructed materials. With tight timelines and very loose student prerequisites, it is easy to lose a portion of the audience. There is a checklist of guidelines that our team has developed through experience. We will oversimplify a few of these guidelines here just to provide a flavor: minimize makefiles, eliminate all possible platform-specific code, avoid language specific idioms, minimize libraries. Some of these are contrary to default practice but prove very effective at keeping the focus on the HPC content. These are continually changing as our experience grows and as our content base diversifies. We look forward to providing this guidance as an open documented resource, and to incorporating the insights of other WAC developers, as discussed in Section V-D.

## 2.5 Hands-On Requirements

*2.5.1 A Central Platform.* The last WAC component described is the hands-on environment itself. We favor very heavy hands-on engagement in all of our agendas for obvious and established pedagogical reasons. To enable this at scale the team must anticipate the worst; otherwise, account and access issues, as well as software environment and system problems, can quickly become catastrophic with hundreds of students. This risk is mitigated by using a central platform with carefully tuned software environments for the students, and real-time monitoring by the WAC staff. With this in hand we can allow every student access to leading-edge hardware with every event. This includes multiple powerful nodes for MPI training, high end CPUs for OpenMP and the latest GPUs for GPU programming or data analytics.

*2.5.2 The Concern with Cloud-based Resources.* The team has found that cloud-based resources are not currently ideal as a hands-on platform. The lack of direct administrator control, combined with and limited access and permission modes, often pushes responsibilities onto the student. This may be acceptable for a months-long or asynchronous course where the student has some time to sort out technical issues and arrive at a functioning configuration. However, this is not workable when expecting students to begin hands-on immediately after the conclusion of the first lecture, and where there are often two or three non-trivial exercises in a single day.

*2.5.3 The Concern with Other Environments.* VMs and docker type environments also tend to push some configuration responsibilities onto the student, and any generic misconfiguration of those images can be disastrous to untangle in real time. Consequently, we prefer a centrally administered resource at the present time. Our team configures these so that the students work in independent interactive partitions, all of which remain readily accessible to the system administrators. We can observe, and correct, any particular problem that any single student complains about. And should we introduce a configuration error ourselves, it can be corrected class-wide without student intervention.

*2.5.4 The Preference for Log on Sites.* All software, both "system" items - compilers, debuggers, software frameworks ("TensorFlow"), etc. - as well as source codes and examples to be utilized in the class, are in place when the students log on. They are never distracted by any installation or configuration issues.

*2.5.5 Monitoring.* Our team is also continually improving our real-time monitoring of the student status and have many possible improvements in the queue. This is critical to quickly eliminate any system-wide issues and can also provide good guidance on when to proceed to the next exercise or lecture. Cloud environments tend to limit the monitoring tools to some specific set with no flexibility. These often aren't applicable to monitoring student progress.

*2.5.6 Flexibility.* Of course, the WAC is agnostic, and any given event can host the exercises on any type of platform. We expect that as cloud resources evolve, many of the above issues will recede and they may eventually become the preferred platform.

## 3 OUR HPC CURRICULUMS

Although the WAC has amplified our training many-fold, our programs remain very much oversubscribed by community demand. Practical HPC programming techniques largely remain hidden within specialized computer science curriculums, and unavailable to scientists. Thus, we endeavor to teach the most useful and demanded content, with an aversion to "flavor of the day" subjects. This "bread-and-butter" approach has also allowed us to refine our materials through iteration, although they are constantly evolving, and new topics appear as HPC progresses and diversifies. Table I includes the four agendas currently in the ACCESS monthly rotation.

TABLE I.  AGENDA OF REGULARLY ROTATING EVENTS.

| Times EST | MPI Day 1 | MPI Day 2 | Big Data Day 1 | Big Data Day 2 | OpenMP | OpenACC |
|---|---|---|---|---|---|---|
| **1100** | Environment | Advanced MPI | Intro to BD | Machine Lrn. | Environment | Environment |
| **1130** | Parallel Comp. | | | | Multicore- | Accelerators |
| **1200** | | | Hadoop | Exercises | Computing | Intro OpenACC |
| **1230** | | | Intro to Spark | Break | Intro OpenMP | |
| **1300** | Break | Break | Break | | Exercises | Break |
| **1330** | | Laplace Soln. | Spark | Deep Learning | Break | Exercises |
| **1400** | MPI Basics | Outro Comp. | | | Exercises | |
| **1430** | | Debuggers | | Exercises | | |
| **1500** | | Profilers | | | Adv. OpenMP | |
| **1530** | Exercises | Exercises | | | | |
| **1600** | Laplace Exer. | | | | OpenMP 4.0+ | CUDA Libs. |
| **1630** | | | Adv. Spark | BD Platforms | | Adv. OpenACC |
| **1700** | Adjourn | Adjourn | Adjourn | Adjourn | Adjourn | Adjourn |

Note the tightly integrated exercises and short days; the latter is important to accommodate multiple time-zones in a humane way. This has become one of our WAC content guidelines. The WAC enables an audience of busy and productive students, faculty, and scientists that would not have the time to travel to a centralized event. We do not squander this by wasting time with filler material but strive to have very tight one- or two-day agendas. These typically last six hours, including breaks, per day. This does require discipline when authoring, with a close eye on the evaluations and other feedback to discern where the pace has been too fast.

Our foremost objective is that students will be able to walk right out of the classroom and apply what they have learned to their own applications. Evaluations show us that this is also their most common objective, and the WAC has been successful in helping them achieve this.

Our latest policy allows accounts to persist for two to three weeks past the conclusion of most events. This is not fully necessary as our authors go to lengths to make sure that our content is not platform specific, and we actively encourage students to immediately apply what they have learned to their home environments. However, some of the newer software stacks, particularly those related to data analytics and machine learning, can be quite delicate. Allowing students to solidify their skills on our working environment, instead of becoming distracted with frustrating software builds, is important. This seems to have the added benefit of encouraging those students to apply for permanent accounts on ACCESS platforms.

As the reader is assumed to be familiar with HPC programming techniques, specifics as regards each of our current rotating curriculums are discussed below.

### 3.1 MPI

MPI remains the centerpiece of scalable computing, and all roadmaps consider the baseline programming environment of the currently planned machines to be "MPI+X". However, MPI retains an intimidating reputation. We dispel that with an agenda designed to get the students writing their own functioning MPI codes from scratch as quickly as possible.

After a brief overview of parallel computing designed to orient them in the landscape of computing paradigms – important as students often bring in prior knowledge of other techniques – we start immediately on basic programs to get them to understand parallel thinking. These build in complexity, but can all be accomplished with a limited selection of MPI routines of the blocking variety.

We then introduce a substantial porting problem: to parallelize a Laplace solver. The serial code has been carefully constructed to be as simple as possible, but it is the same one used for the OpenMP and OpenACC courses. This is critical, as the students can believe that they are not being provided a "rigged" problem to make MPI seem artificially easy. Also, students of those other techniques can compare provided parallel solutions using those techniques to compare the efficiency of the codes. The Laplace solver exercise is scheduled so that students can work on it overnight if they choose. Some students do not require any extra time, and some may choose to work on this non-trivial problem beyond the conclusion of the workshop. Our solution review session is contrived to avoid "spoiling" the exercise for those that require this extra time.

With this experience in hand, we can then provide an overview of the extensive MPI library routines. It is of course impossible to do this thoroughly, and indeed many future MPI programmers will never require most of these routines in their own applications. However, we strive to expose them to the existence and purpose of the various "families" of these routines, and to walk through specific examples for the most important: non-blocking communications, collectives, user-defined datatypes, communicators and single-sided communications. Long-term feedback from previous students has shown this to be an effective balance.

We also briefly expose students to the under-appreciated debugging and profiling tools used in this space. As the most effective tools are proprietary, we do feel an obligation to restrain ourselves lest we violate our portability philosophy.

We conclude with an outgoing overview of alternative software techniques, now that the students have one such paradigm to contrast against. This is useful to dispel any remaining doubts they may have as to the role of MPI in the computing sphere.

Note that exercise times are scattered throughout these presentations, and the students are invited to revisit topics or ask code-related questions at any point. This intermix is far from distracting, but rather keeps the students engaged and leverages the unique WAC environment.

### 3.2 Open MP
Multi-threaded computing is essential to effectively use any modern processor. Recognition of this reality is often slow to come as it requires one to enter the daunting world of parallel computing. Many of our students only reluctantly take this course, but our one-day agenda, promising working knowledge, makes this more palatable, if somewhat hard to believe. Fortunately, the very friendly nature of directive-based programing makes this achievable.

After a brief introduction to the environment and quick overview of modern processors, we jump right into the directives applied to simple loops, and then move quickly along to more complicated loops resembling those they are likely to encounter in their own numerical codes. The students then thread the above-mentioned serial Laplace code to find that they really can obtain good performance results with a few powerful tools.

After they have accomplished something non-trivial, they are exposed to the more sophisticated, if error-prone, parts of OpenMP. While they certainly won't master these less loop-oriented algorithms during the workshop, they will gain an informed appreciation for how flexible and low-level OpenMP can be when necessary.

The last gasps of Moore's Law have instigated a number of design issues that the hardware vendors are hoping to mitigate with new architectures. The latest versions of the OpenMP standard have focused on addressing these, first with the accelerator emphasis of OpenMP 4, and now with the memory hierarchy APIs of the upcoming OpenMP 5. As these are a considerable departure from the previous tack of OpenMP, it would be easy to ignore these in an introductory course. However, it would result in future confusion for the students as they attempt to use the specification, so we provide an overview as our concluding topic.

### 3.3 OpenACC
GPU computing has become an extremely important part of both traditional HPC as well as emerging areas such as machine learning. Although there is hope to use these devices at a high level, such as a library or framework called from Python, the reality for many is that direct programming is required. Previously this meant low-level CUDA, or something similar. The needlessly primitive nature of this approach was discouraging to the knowledgeable and intimidating to the newcomer. Fortunately, just as we were surrendering to the demand to develop some kind of GPU curriculum, OpenACC emerged as the clear alternative. We won't go into history here; suffice it to say that an open standard backed strongly by NVIDIA was a potent combination.

We build upon what was learned with our OpenMP class to enable the same terse, but hands-on, agenda with OpenACC. The two standards bear many similarities – especially their directive-based approach. After a brief introduction to the environment and tools, there is a very shallow overview of GPU computing. This is kept brief as the hardware is rapidly evolving and the students should focus at the more abstract level that OpenACC allows. Indeed, we point out, after their first several successful GPU-utilizing exercises, just how little they actually know about the hardware details.

We again use the same Laplace serial code as our major exercise. This again provides some credibility to our claim that we are not choosing problems contrived to make our techniques seem more effective than they are. Additionally, this provides a very concrete performance comparison against the other paradigms and hardware.

A topic that grew into its own lecture is "Using OpenACC with Libraries". This allows the students to leverage the vast CUDA library ecosystem using only the knowledge that they have acquired that day. The utility of this material is so great that we break it out to provide emphasis and encourage a Q&A session on just this topic.

As with OpenMP, although time is very short, we still wish to expose the students to the more advanced capabilities. They will not have time to use, let alone master, the functionality presented here, but they will be aware of it when their own applications require one of these specialized capabilities.

As per the discussion of Section B, OpenMP has made an effort to support accelerator hardware in a similar manner to OpenACC – they were originally intended to merge. As such we give a very brief overview of this relationship to both avoid future confusion and demonstrate how translatable code is between the two standards.

### 3.4 Hybrid Programming
There is one exceptional event, and topic, that bears special mention as it is rarely attempted outside of a semester-long course: hybrid computing. By this we mean specifically blending MPI, OpenMP and GPU computing so that students can use modern supercomputers effectively (the epitome of "MPI+X" so often seen on HPC roadmaps). This happens during our 4-day Summer Boot Camp, where there is time to work our way through all 3 foundational subjects and culminate with hybrid computing. This is used to launch an optional 3-week hybrid computing challenge contest afterwards. The entrants for this follow-on event consistently demonstrate significant performance gains via parallelization using multiple paradigms. Of course, those that are not successful may choose not to submit an entry, but we find few "dropouts" over the contest window as measured by those that ask to be added to the contest mailing-list at the start versus the final entry count.

### 3.5 Big Data and Machine Learning

This newest addition to our agenda has quickly become the most popular. It is also the most rapidly evolving. The rough outline is to cover big data the first day, and machine learning the second, but we try to build as much as possible on what is taught the first day and not treat them as isolated subjects.

The program starts with an overview of data formats and analysis requirements, which illustrates the shortcomings of traditional approaches. Hadoop gets brief mention as it is unavoidable in this space, although it is quickly superseded by the alternatives that are taught next.

We then progress to Spark, working through many simple examples, but emphasizing scalability and performance continually. There are many short exercises to guarantee that the students can use the tools. The first day ends with the use of Spark clustering libraries as an entry point into the more advanced topics of the second day.

The second day begins using the Spark environment that the students have become comfortable with the first day, but now using very advanced machine learning techniques to develop a recommender system from scratch. The students are provided both small and very large datasets so that they can explore scaling and work with realistic problems.
The curriculum jumps into a very different framework (TensorFlow) for the deep learning portion of the workshop. We develop a character recognition system with several alternative architectures, the last of which is fairly sophisticated and quite up to date in this fast-moving field.

The event concludes with an overview of the very broad big data analytics and machine learning field. This always prompts an extended, user-application focused Q&A, which inevitably runs into our time limit.

Evaluations of this event demonstrate an incredible demand for follow-on material, especially additional neural net architectures. We are considering how we can accommodate this given our current resources. A third day would become logistically difficult for many of our students, who struggle to clear two days of calendar as it is. A separate event would displace one of our other events, barring additional funding.

## 4 RESULTS AND EVALUATION

The WAC has evolved and grown, and there are now significant numbers to report. We believe the sample size for such a controlled and vetted group is unique within HPC.

### 4.1 Demographics

Over the past ten years, the team has delivered 106 events to just over 24,000 students at 90 different sites.

*4.1.1 Geographical Information.* The map of Fig. 1 shows the geographical distribution of sites and the relative student count for events since April of 2015. We are constantly adjusting our balance of new and recurring sites, along with other criteria such as outreach and diversity. The dark-green discs are Established Program to Stimulate Competitive Research (EPSCoR) sites, the green discs are Historically Black Universities (HBUs) and Minority Serving Institutions (MSIs), and the blue discs are both. These are outreach priorities for us. With the WAC enabling dozens of sites per event, we are able to accommodate these competing demands with fine balance.

*4.1.2 Student Profile.* A breakdown by students, shown in Table II, illustrates a diverse array of represented states. Likewise, table III shows the program's recruitment of underrepresented groups, in terms of gender and race. The numbers from these tables reflect a snapshot from the 2016-2017 academic year through the 2021-2022 one and are based on the paper, "A Longitudinal Study of XSEDE HPC Workshops" by Miles, et.al.

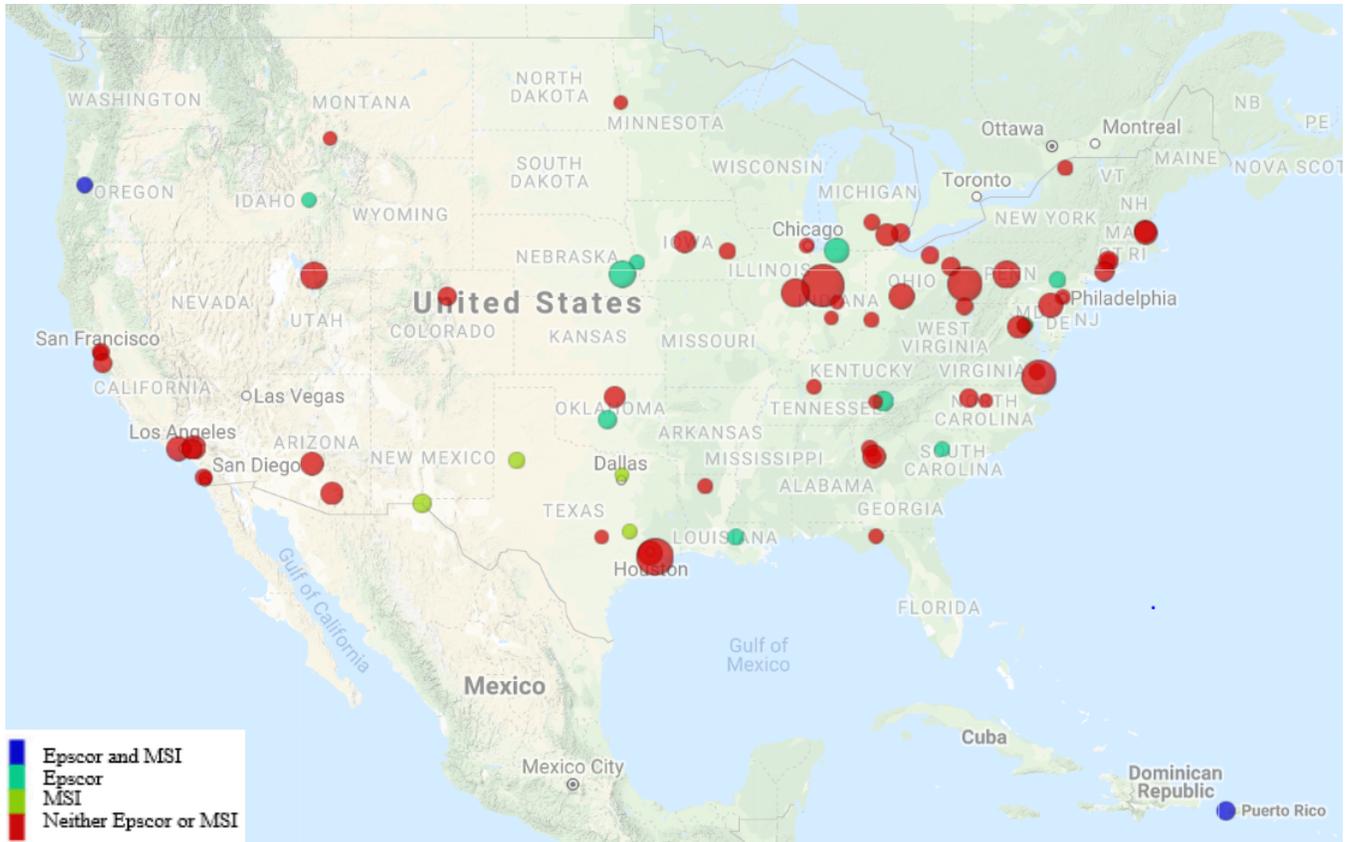

**Figure 1.** Geographical distribution of regularly rotating events since April of 2015.

**Table II:** Workshop Attendance Numbers by Location

| Alabama | 21 | Florida | 38 | Louisiana | 64 |
|---|---|---|---|---|---|
| Alaska | 0 | Georgia | 359 | Maine | 3 |
| Arizona | 331 | Hawaii | 16 | Maryland | 92 |
| Arkansas | 34 | Idaho | 2 | Massachusetts | 403 |
| California | 975 | Illinois | 446 | Michigan | 227 |
| Colorado | 141 | Indiana | 580 | Minnesota | 21 |
| Connecticut | 137 | Iowa | 229 | Mississippi | 21 |
| Delaware | 67 | Kansas | 145 | Missouri | 28 |
| District of Columbia | 212 | Kentucky | 12 | Montana | 1 |

| | | | | | | | | |
|---|---|---|---|---|---|---|---|---|
| Nebraska | 150 | | Oklahoma | 143 | | Utah | 172 |
| Nevada | 31 | | Oregon | 42 | | Vermont | 2 |
| New Hampshire | 0 | | Pennsylvania | 770 | | Virgin Islands | 2 |
| New Jersey | 32 | | Puerto Rico | 60 | | Virginia | 638 |
| New Mexico | 10 | | Rhode Island | 15 | | Washington | 26 |
| New York | 391 | | South Carolina | 24 | | West Virginia | 52 |
| North Carolina | 145 | | South Dakota | 35 | | Wisconsin | 20 |
| North Dakota | 28 | | Tennessee | 137 | | Wyoming | 1 |
| Ohio | 380 | | Texas | 572 | | Outside the US | 145 |

Table III. Self-Reported Gender and Race/Ethnicity Response Count

**Self-Reported Gender**

| | |
|---|---|
| Female | 579 |
| Male | 1954 |
| Other | 22 |
| Prefer not to answer/No response | 879 |

**Self-Reported Race/Ethnicity**

| | |
|---|---|
| American Indian/Alaska Native | 14 |
| Asian | 1144 |
| Black/African American | 117 |
| Hispanic/Latino | 219 |
| Native Hawiian/Other Pacific Islander | 4 |
| Other | 64 |
| White | 1709 |
| No Response | 957 |

### 4.2 Evaluations

Evaluations are requested from the students after every event. They are comprehensive and constantly evolving as we learn which questions to ask. The current survey contains 13 questions, and additionally has several free-form text entry fields. These last often produce the most insightful comments. This data has driven substantial improvements in both the course content and the WAC design. The numbers reflect a high degree of satisfaction. All are well over 4 out of 5. This satisfaction is also reflected at a higher level by the continual backlog of repeat sites as shown in Table IV.

TABLE IV:    Evaluation Data

|  | Big Data | OpenACC | MPI | OpenMP |
|---|---|---|---|---|
| **Respondants** | 445 | 189 | 104 | 170 |
| Survey Question<br>I would recommend this workshop to others. (Out of 5) | 4.31 | 4.49 | 4.43 | 4.31 |
| Survey Question<br>Overall, I would rate the experience as successful. (Out of 5) | 4.19 | 4.38 | 4.33 | 4.26 |

## 5  FUTURE DIRECTIONS

We are confident the WAC model is useful for a wide variety of domains experiencing expertise and instructor shortage, and also for subjects with a lack of critical mass. This means topics where there may only be one, or a few, students at any given institution, but where the national audience is sufficient to justify an event. As such, we hope to proselytize this model wherever possible.

### 5.1  Expansion of Current Programs

The program has a continuous backlog of requests for training, reflected in that every event requires us to select from among the many candidate sites, rejecting others. As noted above, this allows us to tailor demographic targets, but ultimately reflects a failure to satisfy all the student demand.

As we have determined we cannot increase our enrollment ceiling beyond 400 students without compromising the interactive nature of the WAC, we should schedule more events, and are currently looking for resources to do so.

We also would like to expand our course offerings as the feedback discussed in Section IV provides us with several areas of high interest. As this involves the same logistical limits as the general demand issue, we have no immediate ability to do so, but hope that the following initiatives will allow others to do so.

### 5.2  Expansion into Semester Based Curriculums

We are very excited about expanding into semester-based curriculums. At the moment there are discussions with the Pennsylvania State System of Higher Education to deploy the WAC across many of their state schools to enable synchronized weekly classes in a semester format. This would be transformative and allow medium-sized schools to be more ambitious in incorporating HPC content into credit-based courses.

### 5.3  Improvement of the WAC

The WAC is not currently packaged or documented properly for immediate adoption by others. But the team is certain the technology is mature and hopes to do this packaging soon. Finishing the WAC blue book is a key component of this effort.

We also would like to package the tools that we currently use for student monitoring and system configuration into user-friendly versions. A slightly more ambitious goal would be to package the administrative environment, along with the exercises and hands-on tools, as containers such as Dockers or Singularity containers.

## ACKNOWLEDGMENTS

This work used the Extreme Science and Engineering Discovery Environment (XSEDE), which is supported by National Science Foundation grant number OCI-1053575. Specifically, it used the Bridges system, which is supported by NSF award number ACI-1445606, at the Pittsburgh Supercomputing Center (PSC).

Also, this work used the Bridges-2 system at PSC from the Advanced Cyberinfrastructure Coordination Ecosystem: Services & Support (ACCESS) program, which is supported by the National Science Foundation grants #2138259, #2138286, #2137603, and #2138296.

We would like to thank the ACCESS staff, and many sites mentioned herein for making these many workshops possible.